\documentclass[aps, prl, reprint, superscriptaddress, longbibliography]{revtex4-2}

\usepackage{amsmath}
\usepackage{amssymb}
\usepackage{bm}
\usepackage{graphicx}
\usepackage{xcolor}
\usepackage{tikz}

\definecolor{uqpurple}{RGB}{105,39,107}
\definecolor{hotred}{RGB}{210,45,45}
\definecolor{sensorblue}{RGB}{45,105,180}

\newcommand{\dd}{\mathrm{d}}
\newcommand{\Vh}{V_{\mathrm h}}
\newcommand{\Vs}{V_{\mathrm s}}
\newcommand{\Vr}{V_{\mathrm r}}

\newcommand{\Qmax}{\ensuremath{\dot Q_{\scriptscriptstyle
  \mathcal S \subseteq \mathcal{V}_{\mathrm r}}^{\max}}}

\begin{document}

\title{Maximum-Temperature Theorem for Nanothermometry}

\author{Taras Plakhotnik}
\email{taras@physics.uq.edu.au}
\affiliation{
School of Mathematics and Physics,
The University of Queensland,
St Lucia, Queensland 4072, Australia
}

\date{\today}

\begin{abstract}
We derive a sharp upper bound on the   temperature rise of a sensor of arbitrary shape and known thermometric volume $V_\mathrm{s}$, placed anywhere inside a region enclosed by an isothermal reference surface. The bound also requires the maximum net heat power
$\Qmax$ crossing any closed surface
within the region and a lower bound on its thermal
conductivity
$\kappa_{\scriptscriptstyle{\mathcal{V}_{\mathrm r}}}$. It is otherwise independent of the heat-source
distribution, domain shape, and source--sensor distance. For $\Qmax=50~\mathrm{pW}$,
$\kappa_{\scriptscriptstyle{\mathcal{V}_{\mathrm r}}}
=0.1~\mathrm{W\,m^{-1}K^{-1}}$, and
$V_{\mathrm s}$ equal to the volume of a
$60~\mathrm{nm}$-diameter sphere, the theorem permits a maximum
temperature rise of approximately $2~\mathrm{mK}$. Neither source localization nor uncertain source--sensor
separation can therefore account for reported kelvin-scale intracellular
readings under steady-state heat conduction. The result applies to sensors based on 
 nanodiamonds, upconversion nanoparticles, fluorescent polymer beads,
and core--shell quantum dots.\end{abstract}

\maketitle

Temperature is a central but subtle concept in classical thermodynamics defined for equilibrium states but routinely applied
locally to living systems maintained  far from equilibrium. Human body
temperature has been used to identify illness for nearly two centuries and
even served as a reference point for some early thermometer scales.
Measuring temperature inside an individual living cell, however, became
possible only about fifteen years ago. This advance gave rise to an unresolved paradox.

A growing literature now reports intracellular temperatures approximately
\(1\)--\(10~\mathrm{K}\) above the surrounding environment
\cite{Okabe2012,Jiang2017,Chretien2018,Di2022,Xu2025,Zhang2025,Lee2025,Consiglio2026}. These and other examples are discussed in a recent Perspective \cite{Plakhotnik2026}. Such observations
strikingly conflict with estimates based on cellular metabolic power and conventional
heat conduction, which typically predict sub-millikelvin temperature inhomogeneities
on intracellular length scales \cite{Baffou2014,Plakhotnik2026}. Known
experimental artifacts can affect individual probes, but none readily
explains similar results obtained with several physically and chemically different
nanothermometers. Microscopic heat-transport models and molecular-dynamics (MD) simulations have
identified corrections to conventional heat conduction, but these remain
far too small to close the several-orders-of-magnitude gap
\cite{Plakhotnik2026}. 

Most attempts to reconcile the discrepancy  \cite{Kiyonaka2015,Suzuki2015} focus on intracellular heat
sources: their size, shape, internal structure, power distribution,  and
distance from the sensor. These quantities are difficult to determine accurately or verify independently for
complex biological structures such as mitochondria. However,   the temperature of a 
mitochondrion or any other nanostructure is inferred experimentally from a thermosensitive
property of a nearby thermometer. The dimensions and properties of a fabricated nanothermometer are also better
defined and understood than those of cellular heat sources. 

We therefore ask a detector-centered question:\emph{What is the largest  temperature rise
that a sensor of  thermometric volume $V_{\mathrm s}$
can acquire anywhere inside a region $\mathcal V_{\mathrm r}$,
given the relevant  upper bound on  heat power and lower bound on
thermal conductivity?}

\begin{figure}[t]
\centering
\includegraphics[width=0.8\columnwidth]
{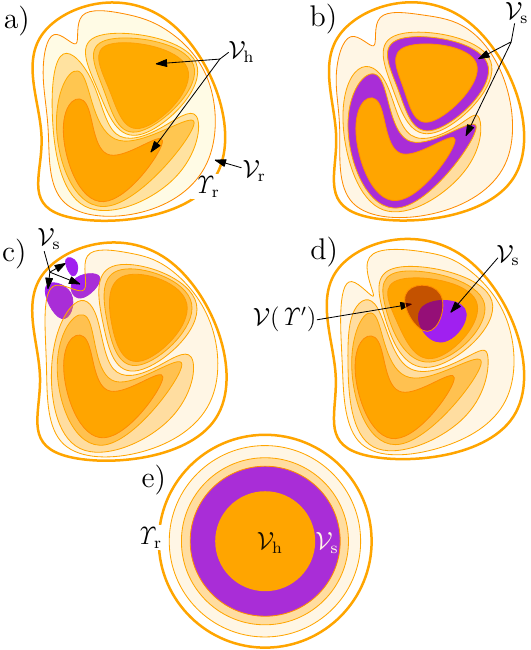}
 \caption{\textbf{Geometrical illustrations of the maximum-temperature theorem.} 
\label{fig:theorem_geometry}
The thin orange curves  illustrate isotherms corresponding to several temperature levels  $\varUpsilon$  and include the influence of the sensor. The thick curves mark the isothermal reference surface \(\mathcal S_{\rm r}\) at temperature \(\varUpsilon_{\rm r}\). Orange and purple shading indicate
$\mathcal V_{\mathrm h}$ and $\mathcal V_{\mathrm s}$, respectively.
 (a) A selected temperature level can enclose several disconnected hot
domains. Their union is denoted by
$\mathcal V_{\mathrm h}$ and its total
volume by $V_{\mathrm h}$. The hot domains expand and may merge as \(\varUpsilon\) decreases.
(b) The largest average temperature  of a sensor of volume
$V_{\mathrm s}$ outside $\mathcal V_{\mathrm h}$ is attained when its
thermometric region $\mathcal V_{\mathrm s}$ occupies the
hottest available volume (not necessarily connected) immediately outside the hot domains. 
(c) Disconnected probes outside the hot domains obey the same bound on their average temperature, with $V_{\rm s}$ equal to their combined thermometric volume.
(d) The sensor may also lie within a hot domain.
The darker region $\mathcal V(\varUpsilon')$, bounded by an
isotherm, has the same volume as the sensor and an average
temperature at least as high.
(e) The bound is approached with uniform conductivity and spherical symmetry.  }
\end{figure}

Temperature reading of a thermometer is determined by the temperature field
throughout the  thermometric volume. 
Let us consider a local temperature field \(T(\mathbf r,t)\). In the following analysis, all quantities are evaluated at the same
instant, and their explicit time dependence is suppressed.
For a given temperature level $\varUpsilon$, all the points satisfying $T(\mathbf r)>\varUpsilon$ form one or several disconnected hot
domains as shown in  Fig.\,\ref{fig:theorem_geometry}(a).  We denote their union by $\mathcal V(\varUpsilon)$ and the boundary of the union by $\mathcal{S}(\varUpsilon)$. The total volume of the union is $V(\varUpsilon)$ and  the area of the boundary  is $A(\varUpsilon)$. We write $\tilde{\mathcal S}\equiv\mathcal S(\varUpsilon)$
for an isothermal boundary, reserving $\mathcal S$ for an
arbitrary closed surface.  
We also assume the existence of a closed reference isothermal surface
$\mathcal S_{\rm r}$ enclosing these
hot domains, on which $T=\varUpsilon_{\rm r}$. The corresponding reference region is
$\mathcal V_{\rm r}=\mathcal V(\varUpsilon_{\rm r})$,
of volume $V_{\rm r}$.  The influence of  the material and heat generation outside \(\mathcal S_{\rm r}\) is contained in the reference temperature \(\varUpsilon_{\rm r}\). The temperature field  is defined by the complete geometry and material properties in $\mathcal{V}_\mathrm{r}$,
including the sensor.

According to Fourier’s heat conduction law, the heat flux $\mathbf J= -\kappa \nabla T $, where $\kappa(\mathbf r)$ is the local heat conductivity which may depend on the position.  Because the boundary $ \mathcal{\tilde{ S}}$ is isothermal, the derivative of $T(\mathbf r)$ along every direction tangent to the boundary is zero, so
$\mathbf J$ is perpendicular to the boundary.   Because an infinitesimal
displacement outward from a hot domain decreases the temperature, 
 the outward normal derivative of $T(\mathbf r)$ is negative. The heat flux
is therefore directed outward from each hot domain.  
The total heat power flowing outward through an isothermal boundary is  
\begin{equation}
\dot Q_{ \mathcal{\tilde{ S}}}= \int_\mathcal{\tilde{ S}}\kappa|\nabla T| \mathrm dS,
\label{eq:Qi_flux}
 \end{equation}

 Using   the integral form of the Cauchy--Schwarz inequality
and Eq.~\eqref{eq:Qi_flux}, 
\begin{equation}\label{eq:ineq_A2}
A^2_ \mathcal{\tilde{ S}}=\left[\int_ \mathcal{\tilde{ S}}
\sqrt{\kappa|\nabla  T|}\frac{1}{\sqrt{\kappa|\nabla T|}}\mathrm dS\right]^2  \le \dot Q_ \mathcal{\tilde{ S}}\int_\mathcal{\tilde{ S}} \frac{\mathrm dS}{\kappa|\nabla T|}.
\end{equation}
 Defining  $\kappa_ \mathcal{\tilde{ S}}$ as the minimum value of $\kappa$ on the isothermal surface and using   inequality~\eqref{eq:ineq_A2}, one  can obtain  
\begin{equation}
A^2_ \mathcal{\tilde{ S}}\kappa_ \mathcal{\tilde{ S}} \le \dot Q_ \mathcal{\tilde{ S}}
\int_ \mathcal{\tilde{ S}}
\frac{\mathrm dS}{|\nabla T|}.
\label{eq:VvsSurfInt}
\end{equation}
The remaining surface integral in Eq.~\eqref{eq:VvsSurfInt} has a direct geometrical interpretation. Consider two neighboring isothermal surfaces corresponding to
temperature levels $\varUpsilon$ and $\varUpsilon+\mathrm d\varUpsilon$.
Let $\mathrm d\ell$ be the magnitude of their local normal separation.  To first order, $\mathrm d\varUpsilon
= |\nabla T|\,\mathrm d\ell$. The small volume between the two surfaces built on a surface element $\mathrm dS$ is 
\begin{equation}
\mathrm dS\,\mathrm d\ell
=\frac{\mathrm dS}{|\nabla T|}\mathrm d\varUpsilon.
\end{equation}

Integrating over the entire isothermal boundary gives the change of the
enclosed volume. Since $\mathrm dV/\mathrm{d}\varUpsilon <0$,
\begin{equation}
\frac{\mathrm d V}{\mathrm d \varUpsilon}
= - \int_{\tilde{\mathcal S}}
\frac{\mathrm dS}{|\nabla T|}.
\label{eq:surf_int}
\end{equation}

The boundary area is not smaller than the area of a sphere of the same volume, \(A\geq(36\pi)^{1/3}V^{2/3}\). Combining this inequality with Eqs.~(3) and (5) yields
\begin{equation}\label{eq:derivative}
-\frac{\mathrm d\varUpsilon}{\mathrm dV} 
 \le \frac{\dot Q_{\tilde{\mathcal{S}}}}{(36\pi)^\frac{2}{3}\kappa_{\tilde{\mathcal S}}V^\frac{4}{3}} ,
\end{equation}
The mathematical approach leading to this equation can be linked to a comparison method for elliptic equations \cite{Talenti1976} and to the heat-flow argument for spherical surface averages in  \cite{Plakhotnik2026}. Here, the use of isothermal surfaces enclosing hotter regions ensures that the normal heat flux is everywhere outward. Equation~\eqref{eq:derivative} will be used below to establish upper bounds on the temperature averaged over a hot domain and over a thermometric volume.  
 
Integration of Eq.~\eqref{eq:derivative} is not straightforward because $\kappa_{\tilde{\mathcal S}}$ and  $\dot Q_{\tilde{\mathcal{S}}}$ may be complex functions of $V$.    To simplify, we consider a steady-state case. Transient
temperatures  are discussed in  Supplemental Material~\cite{SupplementalMaterial}.  

In a steady state, $\dot Q_{\tilde{\mathcal S}}$ equals the
net heat power produced within the region enclosed by
$\tilde{\mathcal S}$. The maximum
net heat power produced within any subregion of
$\mathcal V_{\rm r}$ of volume $V$ is  a $V$-dependent upper bound on $\dot Q_{\tilde{\mathcal S}}$.
To obtain a power bound independent of $V$, let $\Qmax$
be the largest net outward heat power through any closed
surface contained within $\mathcal V_{\rm r}$.
Let $\mathcal V^{+}$ denote the part of the reference region
lying outside the hot region $\mathcal V$, and let
$\kappa_{\mathcal V^{+}}$ be a lower bound on its thermal
conductivity.
For every intervening isotherm,
$\kappa_{\tilde{\mathcal S}}\geq\kappa_{\mathcal V^{+}}$
and $\dot Q_{\tilde{\mathcal S}}\leq\Qmax$.
Integration of Eq.~\eqref{eq:derivative} from $V$ to
$V_{\rm r}$ then gives

\begin{equation}
\varUpsilon-\varUpsilon_{\rm r}
\leq
\frac{3\Qmax}
{(36\pi)^{2/3}\kappa_{\mathcal V^{+}}}
\left(
\frac{1}{V^{1/3}}
-\frac{1}{V_{\rm r}^{1/3}}
\right),
\label{eq:T_surf}
\end{equation}

Because $V(\varUpsilon)$ decreases monotonically, the
inverse function $\varUpsilon(V)$ defines the temperature
of the isothermal boundary enclosing volume $V$.
The hot domain $\mathcal V_{\rm h}$ can then be partitioned
into infinitesimal shells bounded by neighbouring isotherms.
A shell of volume $\mathrm dV$ has temperature
$\varUpsilon(V)$ to first order, and therefore
\begin{equation}
\left\langle T \right\rangle_{\mathcal V_{\mathrm h}}
\equiv
\frac{1}{V_{\mathrm h}}
\int_{\mathcal V_{\mathrm h}}T(\mathbf r)\,\mathrm d^3r
=
\frac{1}{V_{\mathrm h}}
\int_0^{V_{\mathrm h}}\varUpsilon(V)\,\mathrm dV.
\label{eq:Theta_av}
\end{equation}

To bound the average temperature of a hot domain, let
$\kappa_{\scriptscriptstyle{\mathcal{V}_{\mathrm r}}}$
be a lower bound on the thermal conductivity throughout
$\mathcal V_{\rm r}$.
This common bound can be used in Eq.~\eqref{eq:T_surf}
for every nested hot volume.

Combining Eqs.~\eqref{eq:T_surf} and~\eqref{eq:Theta_av}
with the common power and conductivity bounds yields 
\begin{equation}
\left\langle T \right\rangle_{\mathcal V_{\mathrm h}}
-\varUpsilon_{\mathrm r}
\leq
\frac{3\Qmax}
{(36\pi)^{2/3}
\kappa_{\scriptscriptstyle{\mathcal{V}_{\mathrm r}}}}
\left(
\frac{3}{2V_{\mathrm h}^{1/3}}
-\frac{1}{V_{\mathrm r}^{1/3}}
\right).
\label{eq:average_temperature_bound}
\end{equation}

We now consider a nanosensor occupying the thermometric  region
$\mathcal V_{\rm s}$ of volume $V_{\rm s}$, shown in purple in
Fig.~\ref{fig:theorem_geometry}. Throughout the following analysis, the temperature field refers to
the complete configuration, including the sensor and its
thermal properties. 

For a sensor that does not overlap $\mathcal V_{\rm h}$,
the largest average temperature is obtained when its
thermometrically active region occupies the hottest
available volume $V_{\rm s}$ adjacent to the hot domains,
as illustrated in Fig.~\ref{fig:theorem_geometry}(b).
 Consequently,
\begin{align}
&\left\langle T \right\rangle_{\mathcal V_\mathrm{s}} -\varUpsilon_\mathrm{r} 
 \leq
 \frac{1}{\Vs}
 \int_{\Vh}^{\Vh+\Vs} \varUpsilon(V)\,\dd V - \varUpsilon_\mathrm{r}
 \nonumber\\
 &\leq
 \frac{3\Qmax}
 {(36\pi)^\frac{2}{3}\kappa_{\scriptscriptstyle{\mathcal {V}_{\mathrm{r}}}}}
 \left[
 \frac{3}{2}
 \frac{(\Vh+\Vs)^{\frac{2}{3}}-\Vh^{\frac{2}{3}}}{\Vs}
 -\frac{1}{\Vr^{\frac{1}{3}}}
 \right].
 \label{eq:S_shell_average}
\end{align}

The inequality is strict for a sensor separated from the hot domain,
as in Fig.~\ref{fig:theorem_geometry}(c).
Taking the limits $V_{\rm h}/V_{\rm s}\rightarrow0$ and
$V_{\rm r}/V_{\rm s}\rightarrow\infty$ in
Eq.~\eqref{eq:S_shell_average} yields the
\emph{maximum-temperature theorem}:
\begin{equation}
\left\langle T \right\rangle_{\mathcal V_{\mathrm s}}
-\varUpsilon_{\mathrm r}
\leq
\left[\frac{6}{\pi}\right]^{\frac{2}{3}}
\frac{\Qmax}
{8\kappa_{\scriptscriptstyle{\mathcal{V}_{\mathrm r}}}
V_{\mathrm s}^{\frac{1}{3}}}.
\label{eq:sensor}
\end{equation}
A finite $V_{\rm r}$ or a nonzero $V_{\rm h}$
in Eq.~(10) give a tighter bound. The bound is particularly relevant to \mbox{\emph{nanothermometry}}, where
temperature is inferred for a small region within a larger system.

A sensor may also lie within a larger hot domain $\mathcal V(\varUpsilon)$, as shown in
Fig.~\ref{fig:theorem_geometry}(d). In this case, choose $\varUpsilon'\geq \varUpsilon$ so that the volume of
$\mathcal V( \varUpsilon')$ equals the sensor volume. Within the reference region, every point outside
$\mathcal V( \varUpsilon')$ is no hotter than any point inside it. The sensor's average temperature therefore cannot exceed
the average over $\mathcal V( \varUpsilon')$, which is bounded by Eq.~\eqref{eq:average_temperature_bound} with
$V_{\rm h}=V_{\rm s}$. Taking $V_{\rm r}\to\infty$ again gives Eq.~\eqref{eq:sensor}. The theorem therefore requires
no assumption about the position, shape, or connectivity of the thermometric region. Any power converted into heat
by optical excitation or by the probe itself must be included in \Qmax. The concentric spherical construction
in Fig.~\ref{fig:theorem_geometry}(e) attains equality in Eqs.~\eqref{eq:S_shell_average} and \eqref{eq:sensor} as $V_{\rm h}\to0$ and $V_{\rm r}\to\infty$ (see \cite{SupplementalMaterial}). 

The limit \(V_{\rm h}/V_{\rm s}\to\infty\) corresponds to the conventional  situation in which a thermometer is much smaller than the body being measured. The highest steady-state temperature attainable by such a thermometer approaches the surface temperature of the hot body. Indeed, for the same conductivity bound, taking
$V_{\rm s}/V_{\rm h}\to0$ in Eq.~(10)
recovers Eq.~(7). Thus the theorem recovers ordinary contact thermometry when the thermometer is small compared with the measured body. 

The maximum-temperature theorem operates with thermometric volume which is often known much more accurately than the volume of a
biological hot domain and the relevant geometry. For a defect embedded in a nanodiamond, upconversion particle, silica particle, polymer
matrix, or a quantum dot, $V_{\rm s}$ is the host volume sharing the equilibrium 
temperature during the  readout.  For example, using the complete volume  of core/shell   quantum dots as an estimate of \(V_{\rm s}\)  is justified because their  thermalization  within tens of picoseconds \cite{Wieman2026} is much faster than typical thermometric acquisition. For molecular thermometers \cite{Takarada2026} collectively sampling an extended region, \(V_{\rm s}\) is the total volume of all probe molecules contributing to the measurement, rather than the volume of an individual molecule.   

For a numerical example, we take $\Qmax=50~\mathrm{pW}$,
representative of the net heat output measured for cultured mammalian cells
\cite{Dong1989}, $\kappa_{\scriptscriptstyle{\mathcal{V}_{\mathrm{r}}}}=0.1~\mathrm{W\,m^{-1}K^{-1}}$,
close to the lower end of the $0.11$--$0.55~\mathrm{W\,m^{-1}K^{-1}}$
range reported for cells \cite{Plakhotnik2026}, and
$V_{\rm s}=10^{-22}~\mathrm{m^3}$, approximately the volume of a
$60~\mathrm{nm}$-diameter sphere.
Equation~\eqref{eq:sensor} gives an upper bound of approximately
$2~\mathrm{mK}$ on the sensor temperature rise, several orders of
magnitude below the reported intracellular rises of $1$--$10~\mathrm{K}$.
At the same power and conductivity, a rise of $1~\mathrm{K}$ would
formally require an equivalent spherical sensor diameter of about
$0.1~\mathrm{nm}$ or smaller.
For the sensor volume used here, a 1\,K rise requires 
 $\Qmax\gtrsim 25~\mathrm{nW}$, or $\kappa_{\scriptscriptstyle{\mathcal{V}_{\mathrm{r}}}}\lesssim
2\times10^{-4}~\mathrm{W\,m^{-1}K^{-1}}$, or  a suitable combination
of these two factors, even in the ideal arrangement
shown in Fig.~\ref{fig:theorem_geometry}(e).
Other geometries would need lower conductivity or  greater power.

Generally speaking, cell calorimetry measures net heat output, whereas
$\Qmax$ in Eq.~\eqref{eq:sensor}
can be larger if heat generated in one part of the cell
is absorbed elsewhere in the cell.  However, even in the ideal geometry
of Fig.~\ref{fig:theorem_geometry}(e), a $1~\mathrm{K}$ sensor-temperature
rise would invoke  an unrealistic scenario of about 25-nW heat production
and nearly equal absorption elsewhere in the cell. The more realistic
geometry of Fig.~\ref{fig:theorem_geometry}(c) needs even
greater power.

Equation~\eqref{eq:sensor} identifies the volume-averaged local temperature with the thermometer temperature. This requires thermal equilibrium between the thermometrically active degrees of freedom and their local environment. However, different degrees of freedom may equilibrate on widely separated timescales. To illustrate this separation and examine its sensitivity to microscopic force constants, which may be poorly known in complex biomolecular systems, we performed MD simulations of two three-dimensional gas reservoirs separated by a two-dimensional elastic membrane. The model and analysis are given in Supplemental Material~\cite{SupplementalMaterial}.   
\begin{figure}[t]
\centering
\includegraphics[width=0.8\columnwidth]
{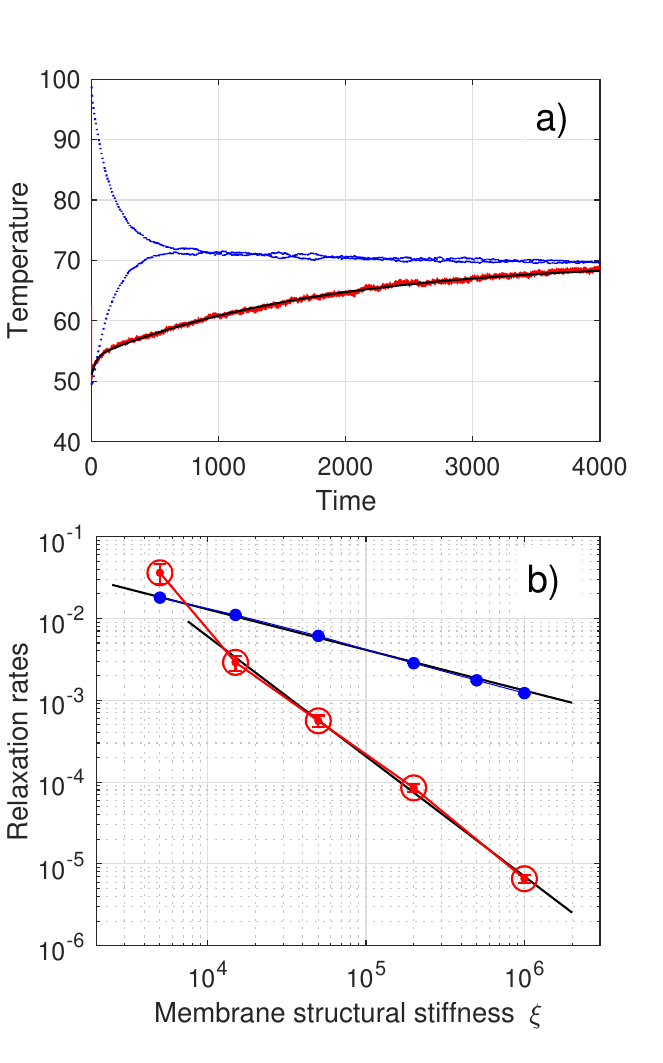}
\caption{\textbf{Relatively fast gas equilibration  and much slower relaxation of the membrane.}
(a) Temperatures of the initially hot and cold gas compartments (dotted blue) and
the separating membrane (red) for structural stiffness
\(\xi=5\times10^{4}\). The smooth black curve is a two-exponential fit to the membrane temperature (see Supplemental Material~\cite{SupplementalMaterial}). The gas temperatures approach each other rapidly,
whereas the membrane continues to relax on a much longer time scale.
(b) Gas-equilibration rate \(\gamma_{\rm g}\) (blue) and  slow membrane-relaxation  rate
\(\gamma_{\rm ms}\) (red, open circles) as functions of \(\xi\). Symbols show simulation
results and black lines are power-law fits. The fitted dependences,
\(\gamma_{\rm g}\propto\xi^{-0.50\pm0.005}\) and
\(\gamma_{\rm ms}\propto\xi^{-1.47\pm0.05}\), show that increasing stiffness suppresses
the slow membrane relaxation rate much more strongly than gas equilibration. The uncertainties are the standard errors of the exponents
obtained from the log--log power-law fits.}
\label{fig:md_rates}
\end{figure}

Figure~\ref{fig:md_rates} shows gas and membrane equilibration
rates versus membrane stiffness $\xi$.
Gas equilibration, mediated by the membrane, depends
comparatively weakly on stiffness,
\(\gamma_{\rm g}\propto\xi^{-0.5}\), whereas the slow membrane
rate decreases much more rapidly,
\(\gamma_{\rm ms}\propto\xi^{-1.5}\).
Thus, for stiff membranes, the gas reservoirs can be close to
mutual equilibrium while the membrane remains out of thermal
equilibrium with them.
Equilibrium calibration does not reveal this separation of timescales.
However, selective energy deposition into slowly relaxing modes
can produce an apparent hot spot if their excitation is interpreted
using equilibrium calibration.
Intracellular fluorescence relaxation on the scale of seconds
after a localized heat pulse has been reported
\cite{Takarada2026}.

The comparatively weak stiffness dependence
of heat transfer within this model reinforces the earlier conclusion
\cite{Plakhotnik2026} that plausible uncertainty in microscopic
force constants is unlikely to account for the discrepancy
of many orders of magnitude between the reported temperature
rises of intracellular nanothermometers and conventional
heat-conduction predictions.

Because the theorem is based on Fourier’s law of heat conduction,  departures from this law could provide sufficient
thermal isolation to explain the reported kelvin-scale readings.
Ballistic transport and quasiballistic size effects (Casimir--Knudsen
corrections) are sometimes invoked for this purpose.
These effects become important when the heat-source dimension is comparable
to the heat-carrier mean free path $\lambda$, notably in crystalline and
low-dimensional solids \cite{Chen2021}. However, a small effective
conductivity assigned to a ballistic layer does not imply strong thermal
insulation. In the ballistic limit, the effective conductivity scales
with layer thickness, leaving the conductance per unit area independent
of thickness and comparable to the Fourier value for a layer of thickness $\lambda$ with bulk thermal conductivity.
In water-rich intracellular environments, where transport becomes
diffusive over molecular distances, substantial corrections of this
kind remain highly speculative \cite{Plakhotnik2026}.

MD simulations of heat transfer between a silver nanoparticle
and water provide a complementary test of continuum heat
conduction \cite{Rajabpour2019}. A model based on the Fourier
heat equation reproduced the main features of the water
temperature evolution, with an inferred thermal conductivity
in the first water shell about \(1.5\) times the bulk
value. This supports using the continuum heat-conduction approximation
at nanometer scales when applying the theorem to water-rich media.

The maximum-temperature theorem replaces poorly known source geometry and source--sensor separation by the measurable thermometric volume of the sensor. This general physical bound provides a practical benchmark for assessing an entire family of nanothermometry techniques, particularly in environments where reliable calibration is difficult and nonthermal factors affect the sensor response. For cellular heat powers, it restricts the steady-state
temperature rise of a typical nanoparticle thermometer to
the millikelvin scale, irrespective of where heat is generated. Kelvin-scale readings must therefore represent either a nonequilibrium sensor state, an unaccounted heat input, or a nonthermal response. Distinguishing these possibilities requires experiments that resolve the relaxation of the thermometer itself, rather than calibration against equilibrium temperatures alone.

\noindent\textit{Data availability---}
The MD source code and the numerical data
supporting the simulation results are available from the author
upon reasonable request.  
\begin{acknowledgments}
ChatGPT and Codex (OpenAI, including GPT-6) were used under the author's direction to assist with coding MD simulations,  checking mathematical arguments, and manuscript preparation. The author reviewed,  revised, and validated  the outputs, verified the final content, and is fully responsible for the work.
\end{acknowledgments}

\bibliography{max_temp_bibl}
 

\clearpage
\onecolumngrid
\setcounter{secnumdepth}{2}
\renewcommand{\thesubsection}{\Alph{subsection}}
\setcounter{subsection}{0}

\setcounter{equation}{0}
\renewcommand{\theequation}{S\arabic{equation}}
\setcounter{figure}{0}
\renewcommand{\thefigure}{S\arabic{figure}}
\setcounter{page}{1}
\begin{center}
{\large\bf Supplemental Material for}\\[1mm]
{\large\bf ``Maximum-Temperature Theorem for Nanothermometry''}

Taras Plakhotnik, School of Mathematics and Physics, The University of Queensland, QLD 4072, Australia
\end{center}

 
\section{Extensions and special cases of the theorem}

\subsection{Sharpness}
In the  example below,
$\dot Q_{\rm r}$ denotes the total heat power generated
within $\mathcal V_{\rm r}$. No heat absorption is included,
so $\dot Q_{\rm r}=\Qmax$. For concentric spherical geometry with constant conductivity
$\kappa_{\scriptscriptstyle{\mathcal {V}_{\mathrm{r}}}}$, the temperature outside a spherical heat-generating
domain of radius \(r_{\rm h}\) is
\begin{equation}
T(r)-\varUpsilon_{\rm r}
=
\frac{\dot Q_{\rm r}}{4\pi\kappa_{\scriptscriptstyle{\mathcal {V}_{\mathrm{r}}}}}
\left(
\frac{1}{r}-\frac{1}{r_{\rm r}}
\right),
\qquad r_{\rm h}\leq r\leq r_{\rm r}.
\label{eq:S_spherical}
\end{equation}
At \(r=r_{\rm h}\), Eq.~\eqref{eq:S_spherical} gives 
equality in Eq.~(7) of the main text. Averaging Eq.~\eqref{eq:S_spherical}
over the concentric sensor shell \(r_{\rm h}<r<r_{\rm s}\), with
\(4\pi(r_{\rm s}^{3}-r_{\rm h}^{3})/3=V_{\rm s}\), gives equality
in Eq.~(10). The bounds are therefore sharp. Because of its simplicity, the spherical configuration  is  commonly used for back-of-the-envelope estimates of hot-region temperatures. 
\subsection{Refinement of the bound}
If heterogeneous conductivity $\kappa(\mathbf{r})$  is known, using
a single global lower bound is often too conservative.    A more precise refinement follows by applying the Cauchy--Schwarz inequality
in the form
\begin{equation}\label{eq:CS_ineq}
\left[\int_{\tilde{\mathcal S}}\sqrt{\kappa(\mathbf r)}\,\mathrm dS\right]^2 = \left[\int_{\tilde{\mathcal S}}\sqrt{\kappa(\mathbf r)|\nabla T|}  \frac{\mathrm dS}{\sqrt{|\nabla T|}}\right]^2
\leq \dot Q_{\tilde{\mathcal S}}\int_{\tilde{\mathcal S}}\frac{\mathrm dS}{|\nabla T|}.
\end{equation}
Thus, \(A^2_{\tilde{\mathcal S}}\kappa_{\tilde{\mathcal S}}\) in Eq.\,(3) of the main text can be replaced by
\(\left[\int_{\tilde{\mathcal S}}\sqrt{\kappa(\mathbf r)}\,\mathrm dS\right]^2\). Determining the actual isotherms generally requires finding the temperature field. Instead, for each volume \(V\), consider all closed surfaces \(\mathcal S_V\) contained within the reference region and enclosing a region of volume \(V\). Define  \(\kappa_{\rm glb}(V)\),  the greatest lower bound over all such surfaces:
\begin{equation}
\kappa_{\rm glb}(V)
\equiv
\frac{1}{(36\pi)^{2/3}V^{4/3}}
\operatorname*{inf}_{\mathcal S_V}
\left[
\int_{\mathcal S_V}\sqrt{\kappa(\mathbf r)}\,\mathrm dS
\right]^2.
\label{eq:S_effective_conductivity}
\end{equation}
Equation~(6) of the main text remains valid with \(\kappa_{\tilde{\mathcal S}}\) replaced by \(\kappa_{\rm glb}(V)\). The isoperimetric inequality ensures that \(\kappa_{\rm glb}(V)\geq\kappa_{\mathcal V_{\rm r}}\), so this construction gives a bound at least as strong as that obtained using the smallest value of \(\kappa\) in $\mathcal{V}_\mathrm{r}$. The volume dependence of \(\kappa_{\rm glb}(V)\) should be retained during the subsequent integration.
This refinement requires knowledge of \(\kappa(\mathbf r)\), but remains independent of the size, location, and geometry of the heat sources. For the direct bound on an isothermal sensor obtained from Eq.~(7), only the isotherms between its boundary and the reference surface are needed. In this case, the minimization can be restricted to surfaces enclosing the entire sensor. Evaluating this refinement can be computationally demanding for a general conductivity field, although simple conductivity distributions may permit analytical or certified numerical estimates.
\providecommand{\Qmaxtime}{%
  \ensuremath{\dot Q_{\tau_{\rm s},
  {\scriptscriptstyle\mathcal F\subseteq\mathcal V_{\rm r}}}^{\max}}}

\subsection{Anisotropic thermal conductivity}
For a symmetric anisotropic conductivity tensor \(\bm K(\mathbf r)\) with smallest eigenvalue within $\mathcal{V}^+$ larger than \(\kappa_{\mathcal V^{+}}\):
 $\mathbf u\cdot\bm K(\mathbf r)\mathbf u
 \geq
 \kappa_{\mathcal V^{+}}|\mathbf u|^2
 \quad
 (\mathbf r\in\mathcal V^{+})$
for every vector \(\mathbf u\). Let \(\mathbf n\) point outward from a hot domain. On an isothermal boundary,
\(\nabla T=-|\nabla T|\mathbf n\). The flux
\(\mathbf J=-\bm K\nabla T\) need not be perpendicular to the boundary, but
its outward component satisfies
\begin{equation}
 J_\perp\equiv\mathbf n\cdot\mathbf J
 =|\nabla T|\,\mathbf n\cdot\bm K\mathbf n
 \geq
 \kappa_{\mathcal V^{+}}|\nabla T|.
 \label{eq:S_normal_flux}
\end{equation}
Energy conservation and the Cauchy--Schwarz inequality then give
\begin{equation}
 A^2_{\tilde{\mathcal S}} \leq
 \left(\int_{\tilde{\mathcal S}}J_\perp\,\mathrm dS\right)
 \left(\int_{\tilde{\mathcal S}}\frac{\mathrm dS}{J_\perp}\right) \leq
 \frac{\dot Q_{\tilde{\mathcal S}}}
 {\kappa_{\mathcal V^{+}}}
 \int_{\tilde{\mathcal S}}\frac{\mathrm dS}{|\nabla T|}.
 \label{eq:S_tensor_CS}
\end{equation}
This replaces Eq.~(2) of the main text; the rest of the derivation is unchanged.

\subsection{Uniform volumetric heat generation}
The maximum-temperature theorem permits heat generation to be
concentrated within an arbitrarily small volume. To illustrate
how conservative the resulting bound can be, consider uniform
volumetric heat generation throughout $\mathcal V_{\rm r}$. The heat power enclosed by isothermal boundaries of
total volume $V$ is then
$\dot Q_{\tilde{\mathcal S}}(V)
=\dot Q_{\rm r}V^{-1}_{\rm r} V.$
Substitution of this expression into Eq.~(6) of the main text, with
$\kappa_{\tilde{\mathcal S}}\geq\kappa_{\scriptscriptstyle{\mathcal {V}_{\mathrm{r}}}}$, and integration between
\(V\) and \(V_{\rm r}\) yield
\begin{equation}
\varUpsilon(V)-\varUpsilon_{\rm r}
\leq
\frac{3\dot Q_{\rm r}}
{2(36\pi)^{\frac{2}{3}}\kappa_{\scriptscriptstyle{\mathcal {V}_{\mathrm{r}}}}V_{\rm r}}
\left(
V_{\rm r}^{\frac{2}{3}}-V^{\frac{2}{3}}
\right).
\label{eq:S_uniform_temperature}
\end{equation}

The highest average temperature of a thermometric volume $V_{\rm s}$ is
obtained by averaging Eq.~\eqref{eq:S_uniform_temperature} over the hottest
volume, $0<V<V_{\rm s}$. This gives
\begin{equation}
\left\langle T\right\rangle_{\mathcal V_{\rm s}}
-\varUpsilon_{\rm r}
\leq
\frac{3\dot Q_{\rm r}}
{2(36\pi)^{\frac{2}{3}}\kappa_{\scriptscriptstyle{\mathcal {V}_{\mathrm{r}}}}V_{\rm r}}
\left(
V_{\rm r}^{\frac{2}{3}}
-\frac{3}{5}V_{\rm s}^{\frac{2}{3}}
\right).
\label{eq:S_uniform_sensor}
\end{equation}
For $\dot Q_{\rm r}=50~\mathrm{pW}$,
$V_{\rm r}=4~\mathrm{pL}$,
$\kappa_{\scriptscriptstyle{\mathcal {V}_{\mathrm{r}}}}=0.1~\mathrm{W\,m^{-1}K^{-1}}$, and
$V_{\rm s}\to 0$,
Eq.~\eqref{eq:S_uniform_sensor} gives
$\left\langle T\right\rangle_{\mathcal V_{\rm s}}
-\varUpsilon_{\rm r}
\lesssim 2.0~\mu\mathrm{K}.$
This is about three orders of magnitude below the
distribution-independent $2~\mathrm{mK}$ maximum obtained in the main text.

\subsection{Bound for time-averaged temperature}
The geometrical argument of the main text also applies to a
time-dependent temperature field $T(\mathbf r,t)$.
For each fixed temperature level $\varUpsilon$, the corresponding
region $\mathcal V(\varUpsilon,t)$, its volume $V(\varUpsilon,t)$,
its isothermal boundary $\mathcal S(\varUpsilon,t)$, the heat power
$\dot Q_{\tilde{\mathcal S}}(t)$, and the conductivity bound
$\kappa_{\tilde{\mathcal S}}(t)$ generally depend on time.
The inverse relation $\varUpsilon(V,t)$ obeys the inequality given in
Eq.~\eqref{eq:derivative}:
\begin{equation}
-\frac{\partial\varUpsilon(V,t)}{\partial V}
\leq
\frac{\dot Q_{\tilde{\mathcal S}}(t)}
{(36\pi)^{2/3}\kappa_{\tilde{\mathcal S}}(t)V^{4/3}}.
\label{eq:derivative_transient}
\end{equation}
The instantaneous heat power $\dot Q_{\tilde{\mathcal S}}(t)$
can be very large over a short interval. We consider the
temperature averaged over an interval $\tau_{\rm s}$, which
may be set by the sensor's temporal resolution or characterise
the duration of a temperature rise. The following bound
applies in either case.

For simplicity, we assume a stationary medium with time-independent
conductivity $\kappa(\mathbf r)$ and a fixed reference region
$\mathcal V_{\rm r}$ whose boundary is held at temperature
$\varUpsilon_{\rm r}$. We define the temperature averaged over
an observation interval $\tau_{\rm s}$ by
\begin{equation}
\overline T(\mathbf r)
=\frac{1}{\tau_{\rm s}}
\int_0^{\tau_{\rm s}}T(\mathbf r,t)\,\mathrm dt.
\label{eq:S_timeavg_fields}
\end{equation}

Let $\mathcal F\subseteq\mathcal V_{\rm r}$ be a fixed spatial
subregion of arbitrary volume $V$ with closed boundary, and let $U_{\mathcal F}(t)$
be the thermodynamic internal energy of its contents. \emph{Over  the observation interval} $\tau_\mathrm{s}$, the change $\Delta U_{\mathcal F}$ is determined by (1) $W_{\mathcal F}^{\rm on}$, the net external work done on
the contents of $\mathcal F$; (2) $ Q_{\mathcal F}$ , the heat supplied to or removed from $\mathcal F$; and (3) $\Delta U_{\mathcal F}^{\rm dif}$, the energy carried into the region by material minus the energy carried out, including flow work (which is excluded from $W_{\mathcal F}^{\rm on}$).  If no material enters or leaves $\mathcal F$, then
$\Delta U_{\mathcal F}^{\rm dif}=0$.

With heat taken positive outward, the first law gives the heat
power averaged over the observation interval:
\begin{equation}
\dot Q_{\tau_{\rm s},
{\scriptscriptstyle\mathcal F\subseteq\mathcal V_{\rm r}}}
\equiv
\frac{Q_{\mathcal F}}
{\tau_{\rm s}}
=
\frac{
W_{\mathcal F}^{\rm on}
+\Delta U_{\mathcal F}^{\rm dif}
-\Delta U_{\mathcal F}}
{\tau_{\rm s}}.
\label{eq:S_timeavg_energy_balance}
\end{equation}

At approximately constant pressure, the energy balance can be
expressed in terms of enthalpy. For chemical heating with reaction
energy thermalised locally, no work other than pressure-volume
work, and negligible thermal transport by material motion, the
reaction and thermal contributions can be separated.
Let $\Delta H_i$ be the molar enthalpy change of reaction $i$
and $\nu_i(\mathbf r,t)$ its net molar reaction rate per unit volume.
The reaction enthalpies and the volumetric heat capacity at constant pressure
$c_{\rm vol}(\mathbf r)$ are evaluated
near the reference conditions and treated as time independent.
\begin{equation}
\dot Q_{\tau_{\rm s},
{\scriptscriptstyle\mathcal F\subseteq\mathcal V_{\rm r}}}
\simeq
 \frac{1}{\tau_{\rm s}}
\left[\int_{\mathcal F}
c_{\rm vol}(\mathbf r)
\bigl[T(\mathbf r,0)-T(\mathbf r,\tau_{\rm s})\bigr]\mathrm d^3r -
\int_0^{\tau_{\rm s}}\int_{\mathcal F}
\sum_i\Delta H_i \nu_i(\mathbf r,t)\mathrm d^3r\,\mathrm dt\right].
\label{eq:S_reaction_heat_balance}
\end{equation}
The first term accounts for the enthalpy change caused by a small
temperature change during the observation interval. The reaction term accounts for enthalpy changes and the associated
heat release or absorption in chemical reactions at constant pressure.

Let $\Qmaxtime$ be the maximum of
$\dot Q_{\tau_{\rm s},
{\scriptscriptstyle\mathcal F\subseteq\mathcal V_{\rm r}}}$
over all such fixed subregions.
Applying the geometrical argument of the main text to the isotherms
of $\overline T$ gives
\begin{equation}
\left\langle\overline T\right\rangle_{\mathcal V_{\rm s}}
-\varUpsilon_{\rm r}
\leq
\left(\frac{6}{\pi}\right)^{2/3}
\frac{\Qmaxtime}
{8\kappa_{\mathcal V_{\rm r}}V_{\rm s}^{1/3}}.
\label{eq:S_timeavg_flux_bound}
\end{equation}
This has the same form as Eq.~\eqref{eq:sensor} in the main text.

The quantity $Q_{\tau_{\rm s}}^{\max}\equiv\tau_{\rm s}\Qmaxtime$ is the largest net heat
transferred out of any fixed subregion during the observation
interval. For $\tau_{\rm s}=1~\mathrm{ms}$, using the conductivity
and thermometric volume of the main-text example, a sensor-temperature
rise of $2~\mathrm K$ averaged over this interval requires
$Q_{\tau_{\rm s}}^{\max}\gtrsim50~\mathrm{pJ}$.
For comparison, a mean heat power of $50~\mathrm{pW}$ corresponds
to $50~\mathrm{fJ}$ over the same interval, approximately
$10^3$ times less. If a chemical process releases heat in pulses much shorter than $\tau_{\rm s}$, this 
does not reduce the required energy. 

Even if the entire mean heating power of $50~\mathrm{pW}$
is assigned to such events, their sustainable average repetition
rate cannot exceed approximately
$50~\mathrm{pW}/(50~\mathrm{pJ})=1~\mathrm{Hz}$.
Release of initially stored thermal energy is included through
Eq.~\eqref{eq:S_timeavg_energy_balance}. However, sustained repetition of heat pulses 
requires replenishing that energy.

\subsection{Estimate of a thermometric volume for molecular probes }
\begin{figure}[htbp]
   \centering
   \includegraphics[width=0.65\columnwidth] {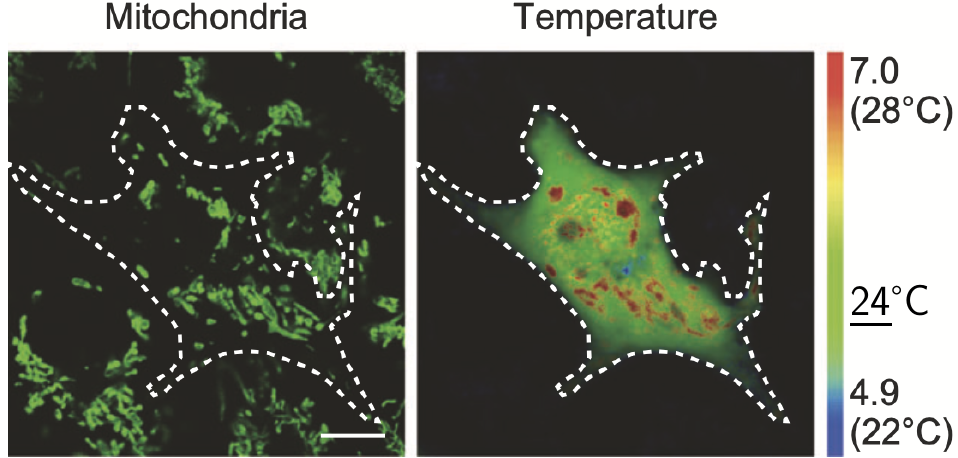} 
   \caption{Localisation of mitochondria on the left (scale bar, 10 \textmu m) and fluorescence lifetime image on the right  from Fig.~1j of Ref.~\cite{Takarada2026}. The cell border is shown by the dashed line. Colour scale shows fluorescence lifetime in ns, with temperatures assigned in the  reference. The green colour in the right panel corresponds approximately
to the culture-medium temperature of $24^\circ\mathrm C$ (tick \& label added here). 
Licensed under Creative Commons Attribution 4.0
(\protect\url{https://creativecommons.org/licenses/by/4.0/}).}
   \label{fig:molecular_probe}
\end{figure}

The total thermometric volume of molecular probes is a bit challenging to estimate but can be obtained from
their loading and spatial distribution. In Ref.~\cite{Takarada2026},
approximately $2~\mathrm{fL}$ of a $0.5\%$ (w/v) fluorescent-polymer
solution was injected per cell, corresponding to a polymer mass
$m_{\rm p}\simeq10^{-14}~\mathrm g$. Taking a polymer density
$\rho_{\rm p}\simeq1~\mathrm{g\,cm^{-3}}$ gives a total probe volume
of approximately $10^{-20}~\mathrm{m^3}$.
The red regions in Fig.~1j of that reference (see also Fig.\,\ref{fig:molecular_probe}) occupy a fraction
$f\simeq0.02$ of the cell's projected area. Assuming  uniform
probe density within the cell, their combined thermometric volume is
\begin{equation}
V_{\rm s}\simeq
f\eta\frac{m_{\rm p}}{\rho_{\rm p}},
\end{equation}
where $\eta$ accounts
for the fraction of the cell thickness sampled along the optical axis.
For full-depth sampling, $\eta\simeq1$.
Taking an axial sampling depth of approximately $1~\mu\mathrm m$
and a cell thickness of approximately $10~\mu\mathrm m$ gives
$\eta\simeq0.1$ and hence $V_{\rm s}\simeq 2\times10^{-23}~\mathrm{m^3}$. The corresponding bound on the temperature rise is about 3.5 mK, using \Qmax  and $\kappa_{\scriptscriptstyle{\mathcal{V}_{\mathrm r}}}$ as in the numerical example of the main text. Since the temperature bound scales as $V_{\rm s}^{-1/3}$,
even a tenfold uncertainty in the estimated thermometric volume
changes the bound by only a factor of approximately two.

\section{Three-dimensional gas--membrane model}
The molecular-dynamics calculation tests whether equilibration of a
surrounding medium implies equilibration of a nanoscale object. It contains
two three-dimensional gas reservoirs separated by an elastic membrane.
Reduced units set the particle mass, \(k_{\rm B}\), and the gas--gas
potential-minimum distance \(r_{\min,g}\) to unity. The physical units depend on the chosen energy, length, and mass scales.
For example, an energy unit of \(5.52\times10^{-23}~\mathrm{J}\),
a length unit of \(5\times10^{-10}~\mathrm{m}\), and a mass unit of
\(5\times10^{-26}~\mathrm{kg}\) give a temperature unit of
\(4~\mathrm{K}\), a time unit of \(15~\mathrm{ps}\), and a stiffness
unit of approximately \(2.2\times10^{-4}~\mathrm{N\,m^{-1}}\).
With this choice, the largest stiffness used, \(\xi=10^6\), corresponds
to approximately \(220~\mathrm{N\,m^{-1}}\), comparable to molecular
bond-stretching force constants, and the equilibrium temperature is
close to room temperature.

The box dimensions are
\(L_x\times L_y\times L_z=124\times31\times31\). The membrane consists of
\(N_y\times N_z=32\times32\) particles initially placed in the \(yz\)-plane
at \(x=L_x/2\), with lattice spacings
\(a_y=L_y/(N_y-1)=1\) and \(a_z=L_z/(N_z-1)=1\). Its perimeter particles are
fixed, leaving \(30\times30\) mobile particles. Writing the position of node
\((i,j)\) as \(\mathbf r_{ij}\), the implemented membrane potential is
\begin{equation}
 U_{\rm mem}=U_{\rm bond}+U_{\rm tether}+U_{\rm bend},
 \label{eq:S_Umem}
\end{equation}
where 
\begin{align}
 U_{\rm bond}
 &=\frac{k_{\rm bond}}{2}
 \sum_{\langle pq\rangle}
 \left(|\mathbf r_p-\mathbf r_q|-a_{pq}\right)^2,
 \label{eq:S_Ubond}\\
 U_{\rm tether}
 &=\frac{k_{\rm tether}}{2}
 \sum_{i,j}\left(x_{ij}-\frac{L_x}{2}\right)^2,
 \label{eq:S_Utether}
\end{align}

Defining the discrete second differences
$\delta_y^2\mathbf r_{ij}
=\mathbf r_{i+1,j}-2\mathbf r_{ij}+\mathbf r_{i-1,j}$ and
$\delta_z^2\mathbf r_{ij}
=\mathbf r_{i,j+1}-2\mathbf r_{ij}+\mathbf r_{i,j-1},$
the bending energy is
\begin{equation}
U_{\rm bend}
=\frac{k_{\rm bend}}{2}
\left[
\sum_{i=1}^{N_y-2}\sum_{j=0}^{N_z-1}
|\delta_y^2\mathbf r_{ij}|^2
+
\sum_{i=0}^{N_y-1}\sum_{j=1}^{N_z-2}
|\delta_z^2\mathbf r_{ij}|^2
\right].
\label{eq:S_Ubend}
\end{equation}

The bond sum contains nearest neighbours along the \(y\) and \(z\)
directions, with \(a_{pq}=a_y\) or \(a_z\). The tether acts only on the
coordinate normal to the initial membrane plane; it confines the membrane
near \(x=L_x/2\) but is not a thermal reservoir.  In the stiffness series,
\begin{equation}
 \xi=k_{\rm bond}=k_{\rm bend},
 \qquad k_{\rm tether}=15,
 \label{eq:S_stiffness}
\end{equation}
so that only the elastic stiffness was varied.

Gas--gas, gas--membrane, and gas--wall interactions use the
soft-core Weeks--Chandler--Andersen potential
\begin{align}
r_{\rm eff}
&=\left[r^2+(\alpha\sigma_\beta)^2\right]^{1/2},
\qquad \alpha=0.5,
\qquad \beta\in\{\mathrm{g,m,w}\},
\label{eq:S_reff}\\
U_\beta(r)
&=
\begin{cases}
4\epsilon_\beta
\left[
\left(\dfrac{\sigma_\beta}{r_{\rm eff}}\right)^{12}
-\left(\dfrac{\sigma_\beta}{r_{\rm eff}}\right)^6
\right]+\epsilon_\beta,
&r_{\rm eff}<2^{1/6}\sigma_\beta,\\[1mm]
0,
&r_{\rm eff}\geq2^{1/6}\sigma_\beta .
\end{cases}
\label{eq:S_WCA}
\end{align}
Here \(\beta\) specifies whether a gas particle interacts with another gas particle, a membrane particle, or the rigid wall. The interaction parameters are defined as
\[
\epsilon_\mathrm{g}=0.25,\; r_{\min,\mathrm{g}}=1,\;
\epsilon_\mathrm{m}=40,\; r_{\min,\mathrm{m}}=1.5,\epsilon_\mathrm{w}=2,\text{ and } \; r_{\min,\mathrm{w}}=1.
\]
where \(r_{\min,\beta} = \sigma_\beta2^{1/6}\) equals the cutoff in the effective separation for the corresponding  interaction. The equations of motion were integrated using the velocity-Verlet
algorithm. For each value of \(\xi\), eight simulations were performed
with different random seeds for the initial conditions. Time-step
convergence was tested at \(\xi=10^6\) using four seeds for each of the
time steps \(\Delta t=5\times10^{-5}\) and \(2\times10^{-5}\), with
\(4\times10^8\) and \(10^9\) integration steps per run, respectively. No detectable differences in the
temperature trajectories or fitted relaxation rates were found. Therefore  \(\Delta t=5\times10^{-5}\) and \(1.5\times10^9\) steps were used  in the production run.

For each of the subsystems (the initially cold and hot gases and
the membrane) the instantaneous kinetic temperature
\begin{equation}
T_\chi(t)\approx\frac{2K_\chi(t)}
{3N_\chi k_{\rm B}},
\qquad
\chi\in\{\mathrm c,\mathrm h,\mathrm m\},
\label{eq:S_temperature}
\end{equation}
was determined, where \(K_\chi\) is the kinetic energy calculated relative to the
centre-of-mass motion of subsystem \(\chi\), and \(N_\chi\) is the number
of particles in that subsystem. In these simulations, \(N_{\rm c}=N_{\rm h} =4500\), and
\(N_{\mathrm{m}}=900\). Maxwellian velocities were chosen to set the initial kinetic temperatures to \(T_{\rm c}=50\), \(T_{\rm h}=100\), and \(T_{\rm m}=100\).

The initially flat membrane has negligible elastic energy. The rapid initial redistribution between kinetic and potential energies is excluded from the relaxation fits. After this transient, the membrane temperature is fitted by
\begin{equation}
T_{\rm m}(t)
=T_\infty-\Delta T_{\rm mf}\exp(-\gamma_{\rm mf}t)
-\Delta T_{\rm ms}\exp(-\gamma_{\rm ms}t).
\label{eq:S_membrane_fit}
\end{equation}
The common equilibrium temperature \(T_\infty\) is fixed from energy conservation:
\begin{equation}
T_\infty=
\frac{\langle N_{\rm c}T_{\rm c}(t)+N_{\rm h}T_{\rm h}(t)
+2N_{\rm m}T_{\rm m}(t)\rangle}
{N_{\rm c}+N_{\rm h}+2N_{\rm m}}.
\label{eq:S_equilibrium_temperature}
\end{equation}
The factor of two accounts for the equal mean kinetic and elastic energies of the membrane after the initial transient.

The slow term accounts for most of the membrane-temperature change over the fitted interval and defines the global membrane-equilibration rate \(\gamma_{\rm ms}\), used in Fig.~2(b). The fast-component amplitude \(\Delta T_{\rm mf}\) decreases with increasing membrane stiffness. At \(\xi=10^6\), it is approximately one temperature unit, making \(\gamma_{\rm mf}\) difficult to determine reliably.

The gas-equilibration rate \(\gamma_{\rm g}\) is obtained by fitting the temperature difference between the two reservoirs:
\begin{equation}
T_{\rm h}(t)-T_{\rm c}(t)
=2\Delta T_{\rm g}\exp(-\gamma_{\rm g}t).
\label{eq:S_gas_fit}
\end{equation}
Taking the difference eliminates the common drift of the gas temperatures associated with membrane heating, allowing \(\gamma_{\rm g}\) to be determined even when the gas and membrane relaxation rates are comparable.
\end{document}